\documentclass[a4paper]{revtex4-1}
\usepackage{graphicx}
\usepackage{epstopdf,gensymb}
\begin{document}
\title{New mechanisms for generating\\super-ponderomotive electrons in laser-irradiated targets}

\author{A. V. Arefiev$^1$, A. P. L. Robinson$^2$, V. N. Khudik$^1$, B. N. Breizman$^1$, and M. Schollmeier$^3$}
\affiliation{$^1$ Institute for Fusion Studies, The University of Texas, Austin,Texas 78712, USA\\$^2$ Central Laser Facility, STFC Rutherford-Appleton Laboratory, Didcot, OX11 0QX, UK\\$^3$ Sandia National Laboratories, Albuquerque, New Mexico 87185, USA}

\date{\today}

%\author{Alexey Arefiev}
%\address{Institute for Fusion Studies, The University of Texas, Austin,Texas 78712, USA}
%
%\author{A. P. L. Robinson}
%\address{Central Laser Facility, STFC Rutherford-Appleton Laboratory, Didcot, OX11 0QX, UK}
%
%\author{Vladimir Khudik}
%\address{Institute for Fusion Studies, The University of Texas, Austin,Texas 78712, USA}
%
%\author{Boris Breizman}
%\address{Institute for Fusion Studies, The University of Texas, Austin,Texas 78712, USA}
%
%\author{Marius Schollmeier}
%\address{Sandia National Laboratories, Albuquerque, New Mexico 87185, USA}

\begin{abstract}
It is shown that static longitudinal and transverse electric fields can significantly alter electron acceleration by a long laser beam in a sub-critical plasma, enabling generation of super-ponderomotive electrons. The role of the plasma fields in this regime is not to directly transfer substantial energy to the electron, but rather to reduce the axial dephasing rate between the electron and the laser beam. The reduced dephasing in both cases leads to a subsequent enhancement of the axial momentum and total electron energy. These mechanisms can be relevant to experiments with solid-density targets where a sub-critical plasma layer occurs as a result of a considerable prepulse.
\end{abstract}

\maketitle

\section{Introduction}

Electron heating in laser-irradiated targets is crucial for production of energetic ions~\cite{Fuchs2006,Flippo2010} and for other applications, including x-ray generation~\cite{Kneip2008,Cipiccia2011}, positron production~\cite{Chen2013}, and fast ignition~\cite{Tabak1994}.  The number of hot electrons and their maximum energy depend on the pulse duration and on the target density. We focus on electron acceleration in a sub-critical plasma irradiated by a laser pulse that is significantly longer than the period of plasma oscillations. Sub-critical plasmas can naturally occur in experiments with solid-density targets as a result of a prepulse and, therefore, the regime under consideration is not just limited to experiments with rarefied targets like gas jets. We show that quasi-static transverse and longitudinal electric fields that naturally arise in the plasma can lead to a significant electron energy gain, with the maximum electron energy exceeding the ponderomotive energy.

\section{Plasma channeling} \label{Sec_chan}

The key feature of the considered regime is formation of a plasma channel that slowly evolves on an ion time scale~\cite{Pukhov1999,Willingale2013}. The laser beam expels electrons radially, creating a plasma configuration where the radial component of the ponderomotive force acting on plasma electrons is balanced by the static electric field generated by the ions in the channel. The channel is overall neutral and it consists of a positively charged cylinder and a negatively charged shell formed by the expelled electrons. The coaxial structure of the channel generates not only a transverse electric field, but also an axial electric field pointing out of the channel opening. 

Attached animations illustrate the channel formation observed in a 2D PIC simulation using PSC~\cite{PSC}. In this simulation, a laser pulse was normally incident an underdense hydrogenic plasma with density $n_0 = 8\times$10$^{25}$m$^{-3}$. The length and width of the slab were 200$\mu$m and 160$\mu$m, and the simulation domain with open boundaries was 300$\mu$m by 200$\mu$m (12000 by 2000 cells). The laser pulse had $a_0 =10$, $\lambda = 1 \mu$m, a FWHM width of 8$\mu$m, and a duration of 500~fs. All plotted quantities are time-averaged over 7 wave periods. The fields are normalized to $E_0 \equiv 2 \pi a_0 m_e c^2 / |e| \lambda \approx 3.2 \times 10^{13}$ V/m, where $c$ is the speed of light and $m_e$ and $e$ are the electron mass and charge. Denoting the two axes of the simulation domain as $z$ (laser direction) and $y$, the electric field of the laser was polarized in the $x$ direction. A quasi-steady-state pattern is established at each axial location after several plasma periods have passed since the beginning of the interaction with the laser. Similarly to the results reported in Ref.~\cite{Arefiev2012b}, we observe that copious energetic electrons are generated inside the channel following its formation. The energies of the energetic electrons considerably exceed the ponderomotive energy, which is the upper limit on electron energy for a single electron accelerated in a vacuum by a plane wave. For $a_0 = 10$, the ponderomotive energy is $\varepsilon_* \approx a_0^2 m_e c^2 / 2 \approx 25$ MeV. The fact that there are electrons with energies exceeding $\varepsilon_*$ points to the fact that the fields generated by the plasma have a profound effect on electron acceleration despite that the plasma is significantly underdense. In sections \ref{Sec_perp} and \ref{Sec_ax}, we explain how the transverse and axial steady-state fields affect the electron acceleration.

\section{Dephasing and longitudinal acceleration}

At relativistic laser amplitudes, $a_0 \gg 1$, most of the energy of an electron accelerated by a plane wave in a vacuum is associated with its longitudinal motion. This is also the case for the underdense channel under consideration. In order to understand the underlying cause for the enhancement of the axial acceleration, let us consider a two-dimensional setup similar to that used in the simulation, such that the laser electric field is perpendicular to the static fields produced by the channel. The longitudinal momentum balance equation for a single electron in an ion channel according to Refs.~\cite{Arefiev2012b} and \cite{Robinson2013} then has the form
\begin{equation} \label{axial_eq}
\frac{d}{d \tau} \left( \frac{p_z}{m_e c} \right) = \frac{d}{d \xi} \left( \frac{a^2}{2} \right) - \frac{\gamma |e| E_z}{\omega m_e c},
\end{equation}
where $p_z$ is the longitudinal momentum and $E_z$ is the longitudinal electric field in the channel. The normalized wave amplitude $a$ depends only on the phase variable $\xi \equiv \omega (t - z/c)$, where $t$ is the time in the ion frame of reference, $z$ is the distance along the channel, and $\omega$ is the wave frequency. We consider a wave with a slowly varying envelope $a_* (\xi) \leq a_0$, so that $a = a_* (\xi) \sin(\xi)$. In Eq.~(\ref{axial_eq}), $\tau$ is a dimensionless proper time defined by the relation $d \tau / dt = \omega/\gamma$, where $\gamma$ is 
the relativistic factor $\gamma = \sqrt{1 + p^2 / m_e c^2}$ determined by the total electron momentum $p$. 

Equation (\ref{axial_eq}) shows that the longitudinal acceleration of electrons is caused by a gradient of the wave amplitude and by the axial electric field. We consider the case where $E_z$ is relatively weak, so that the momentum gain directly from acceleration by this field is relatively small compared to $m_e c a_0^2$, which is the maximum axial momentum of a free electron accelerated in a vacuum. The main contribution in this case comes from the axial acceleration by the wave through the Lorentz force. The corresponding force oscillates on the right-hand side of Eq.~(\ref{axial_eq}) at twice the wave frequency. Its sign changes every time $\xi$ increases by $\pi$, which means that the resulting acceleration is limited by how fast $\xi$ changes with $\tau$. The phase $\xi$ increases continuously because the electron is moving slower than the wave in the axial direction. The corresponding dephasing rate is $d \xi / d \tau =  \gamma - p_z/m_e c$ and it can be show that it changes with time at the rate 
\begin{equation} \label{deph}
\frac{d}{d \xi} \left( \gamma - p_z/m_e c \right) = - \frac{|e| E_y}{m_e c \omega} \frac{\omega}{c} \frac{d y}{d \tau} + \frac{|e| E_z}{m_e c \omega},
\end{equation}
where $E_y$ is the transverse electric field of the channel. This expression suggests that there exist two ways to reduce the dephasing rate and thus enhance the longitudinal acceleration. The first term on the right-hand indicates that amplification of transverse oscillations can be beneficial, whereas the second term indicates that even a relatively weak axial field can cause the desired reduction as well.

%******************************************************************************************

\section{Role of the transverse static field} \label{Sec_perp}

Here we only give a qualitative explanation of the effect that has been overlooked until recently. It is discussed in detail in Refs.~\cite{Arefiev2012}, \cite{Arefiev2014}, and \cite{Arefiev2012b}. Let us consider an electron in a straight channel similar to that shown in Sec.~\ref{Sec_chan}. For simplicity, the laser electric field (directed along the $x$-axis) is perpendicular to the field generated by the channel (directed along the $y$-axis), so that the laser is unable to directly drive betatron oscillations (oscillations across the channel). Betatron oscillations are then described by an equation, $d^2 y / d \tau^2 + \gamma \omega_p^2 / \omega^2 y = 0$, similar to that of an oscillator with a modulated natural frequency, where $\omega_p$ is the plasma frequency. A significant modulation of the $\gamma$-factor is induced by the axial motion of the electron in the plane wave. The electron is pushed primarily forward experiencing alternating acceleration and deceleration caused by axial gradients of the wave amplitude. 

The electron axial motion is similar to that in a vacuum when the betatron oscillations remain small, with the dephasing rate $\gamma - p_z/m_e c = 1$ and, as a result, $\tau = \xi$. We can therefore 
estimate the natural frequency as $\sqrt{\gamma} \omega_p / \omega \approx a_0 \omega_p / \sqrt{2} \omega$, because $\gamma \approx a_0^2/2$ for a wave of ultra-relativistic amplitude ($a_0 \gg 1$). The $\gamma$-factor changes in time as $\sin^2(\xi)$ for a wave with $a = a_0 \sin(\xi)$, which means that the frequency of the modulations is 2.  

Natural oscillations in this system are stable if their frequency is considerably less than the frequency of the modulations. There is a frequency threshold that is roughly comparable to the frequency of the modulations, $a_0 \omega_p / \sqrt{2} \omega \approx 2$, above which the oscillations become parametrically unstable and their amplitude grows exponentially. The resulting amplification of the betatron oscillations reduces the dephasing rate, as predicted in the previous section. The reduced dephasing leads to enhancement of the axial momentum and thus total electron energy (see Ref.~\cite{Arefiev2012} and \cite{Arefiev2014}). It is important to point out that the amplification condition can be satisfied in a significantly sub-critical plasma for $a_0 \gg 1$.

%An example of electron motion in this regime is shown in Fig.~\ref{fig:traj} (upper panel), where the trajectory is a solution of Eq.~(\ref{main1_2}) without the driving force on the right-hand side. The color-coding represents $| \gamma (\omega_p^2 / \omega^2) y |$ as a function of $\xi$ and $y$. At the electron location, this quantity is proportional to the restoring force acting on the electron.

%Another way of looking at this is by noting that a static transverse electric field is enhanced by a factor of $\gamma$ in the instantaneous reference frame co-moving with the electron along the axis of the channel. 

%******************************************************************************************

\section{Role of the axial static field} \label{Sec_ax}

\begin{figure}[h]
\begin{minipage}{8.5pc}
\includegraphics[width=8.5pc]{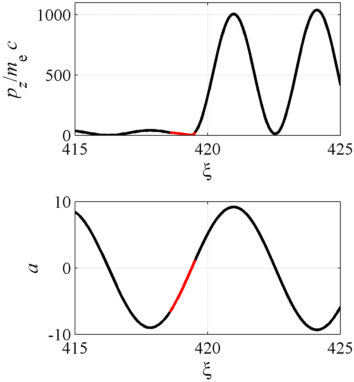}
\end{minipage}\hspace{1pc}%
\begin{minipage}{8.5pc}
\includegraphics[width=8.5pc]{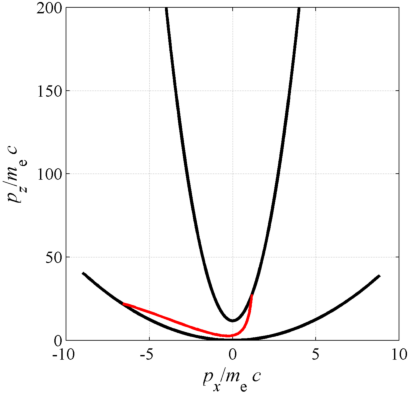}
\end{minipage}\hspace{1pc}%
\begin{minipage}{8.5pc}
\includegraphics[width=8.5pc]{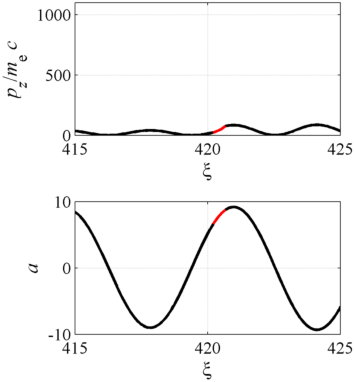}
\end{minipage}\hspace{1pc}%
\begin{minipage}{8.5pc}
\includegraphics[width=8.5pc]{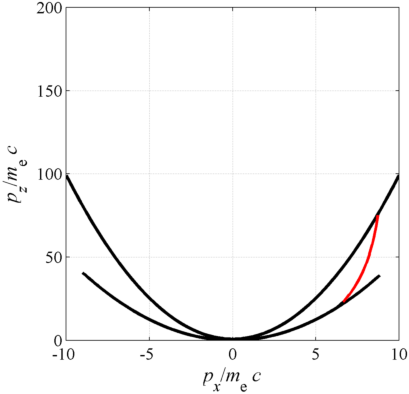}
\end{minipage}
\caption{\label{1D_Ez} Electron axial momentum, the field sampled by the electron, and electron momentum-space for two cases with the axial field located at $142 \mu$m$ \ge z \ge 147 \mu$m (left two panels) and at $144 \mu$m$ \ge z \ge 149 \mu$m (right two panels).}
\end{figure}

The role of the axial static field has been considered in Ref.~\cite{Robinson2013}. Here we wish to draw particular attention to the importance of the timing of the longitudinal boost in this mechanism. The effect can be illustrated by considering what happens after a free electron that is being accelerated by a plane wave crosses a region with a weak axial electric field. The equations that have to be solved in this case are essentially the axial momentum balance equation and the equation for the dephasing rate~\cite{Robinson2013}. We have solved them numerically for a laser pulse with $\lambda = $1 $\mu$m and $a_x = a_0\cos(\xi) \exp\left[-(z-ct-z_0)^2 / 2c^2 t_L^2 \right]$, where $a_0 = 10$, , $z_0 =6ct_L$, and $t_L =$40~fs.  The electron is initially at rest at $z = 0$ $\mu$m. We apply a constant longitudinal electric field $E_z = -0.1 E_L$ over 5 $\mu$m. We consider two cases with the region located at $142 \mu$m$ \ge z \ge 147 \mu$m and at $144 \mu$m$ \ge z \ge 149 \mu$m. The results are shown in Fig.~\ref{1D_Ez}, where the red segments mark electron dynamics during the interaction with the field. 

In the first case, the interaction with the axial field reduces the dephasing rate, $\gamma - p_z/m_e c$, from 1 to 0.04. The maximum axial momentum after the interaction increases by a factor of 25. The plot of $p_z$ vs. $p_x$ confirms that the change of the axial momentum during the interaction is small. The key role of the weak axial field is to launch the electron onto a super-ponderomotive trajectory represented by the steep upper parabola. The second case illustrates the importance of the wave phase $\xi$ during the interaction. In contrast with the first case, $a$ increases as the electron enters the region with the electric field. Both $p_z$ and $\gamma$ are then higher during the interaction, which shortens the time $\Delta \tau$ that it takes for the electron to traverse the interaction region. The corresponding $\Delta \xi$ is also smaller and, in agreement with Eq.~(\ref{deph}), the dephasing rate decreases only to 0.51. This significantly reduces the axial momentum gain following the interaction.

%The equations that describe the electron dynamics in this case are~\cite{Robinson2013}
%\begin{eqnarray}
%&& \frac{d}{dt} \left( \frac{p_z}{m_e c} \right)= - \frac{|e| E_z}{m_e c} + \frac{\omega}{\gamma} \frac{d}{d \xi} \left( \frac{a^2}{2} \right), \\
%&& \frac{d \xi}{d t} = \frac{\omega}{\gamma} \left( \gamma - \frac{p_z}{m_e c} \right),
%\end{eqnarray}
%with the transverse components of the electron momentum given by $p_x = a m_e c$ and $p_y = 0$. The laser electric field is directed along the $x$-axis and propagates along the $z$-axis. Its normalized vector potential is $a = a_* (\xi) \sin(\xi)$, where $a_*$ is a slowly varying envelope. The $\gamma$-factor is $\gamma = \sqrt{1 + a^2 + p_z^2 / m_e^2 c^2}$.

%******************************************************************************************

\section{Summary}

It has been shown that static longitudinal and transverse electric fields can significantly alter electron acceleration by a long laser beam in a sub-critical plasma, enabling generation of super-ponderomotive electrons. The role of the plasma fields in this regime is not to directly transfer substantial energy to the electron, but rather to reduce the axial dephasing rate between the electron and the laser beam. The axial field directly reduces the dephasing, whereas the mechanism is more complex in the case of the transverse field. Axial motion in a plane wave induces a significant modulation in the $\gamma$-factor. This modulation makes betatron oscillations across the channel unstable. The modulations of $\gamma$ effectively modulate the restoring force generated by the ions via the relativistic mass effect. The resulting amplification of the oscillations then reduces the dephasing. The reduced dephasing in both cases leads to a subsequent enhancement of the axial momentum and thus total electron energy.

%******************************************************************************************

\section{Acknowledgments}

AVA was supported by U.S. DoE Contract No. DE-FG02-04ER54742 and by NNSA Contract No. DE-FC52-08NA28512. AA acknowledges the Texas Advanced Computing Center for providing HPC resources. APLR is grateful for computing resources provided by STFC's e-Science facility.  

%******************************************************************************************

%\section*{References}

\end{document}